\documentclass[prd,twocolumn,superscriptaddress,preprintnumbers,nofootinbib,noshowpacs]{revtex4} 
\usepackage{epsfig}
\usepackage{color}
\usepackage{amsmath}
\usepackage{amssymb}
\usepackage{natbib}
\usepackage{enumerate}

\usepackage{graphicx}
\graphicspath{{figs/}}

\def\nn{\nonumber}
\newcommand{\be}{\begin{equation}}
\newcommand{\ee}{\end{equation}}
\newcommand{\beq}{\begin{equation}}
\newcommand{\eeq}{\end{equation}}
\newcommand{\bea}{\begin{eqnarray}}
\newcommand{\eea}{\end{eqnarray}}
\newcommand{\beaa}{\begin{eqnarray*}}
\newcommand{\eeaa}{\end{eqnarray*}}
\newcommand{\ba}{\begin{array}}
\newcommand{\ea}{\end{array}}
\newcommand{\bi}{\begin{itemize}}
\newcommand{\ei}{\end{itemize}}
\newcommand{\ben}{\begin{enumerate}}
\newcommand{\een}{\end{enumerate}}


\begin{document}
\preprint{ULB-TH/15-06}
\title{New Search for Monochromatic Neutrinos from Dark Matter Decay}

\author{Cha\"imae El Aisati}
\email{celaisat@ulb.ac.be}
\affiliation{Service de Physique Th\'eorique - Universit\'e Libre de Bruxelles, Boulevard du Triomphe, CP225, 1050 Brussels, Belgium}
\author{Michael Gustafsson}
\email{michael.gustafsson@theorie.physik.uni-goettingen.de}
\affiliation{Institute for theoretical Physics - Faculty of Physics, Georg-August University G\"ottingen, Friedrich-Hund-Platz 1, D-37077 G\"ottingen, Germany} 
\author{Thomas Hambye}
\email{thambye@ulb.ac.be}
\affiliation{Service de Physique Th\'eorique - Universit\'e Libre de Bruxelles, Boulevard du Triomphe, CP225, 1050 Brussels, Belgium}

\date{\today}

\begin{abstract}
From data recently reported from the IceCube telescope, we derive new bounds on the monochromatic neutrino signal produced from dark matter particle decays. In the few TeV to tens of TeV energy range, these bounds turn out to be better than previous limits by more than an order of magnitude. As a result,  intensity constraints on neutrino lines at energies above a few TeV are now comparable to those on gamma-ray lines. From the same data sample, we also perform a detailed search for a neutrino line, showing that there is no significant hint for such a signal. 
\end{abstract}

\maketitle

\section{Introduction}
The observation of a high-energy spectral line in the cosmic gamma-ray spectrum is one of the cleanest indirect detection signatures one could think of for establishing the existence of a dark matter (DM) particle. Such a line could be produced from DM particle annihilations or decays in our galaxy \cite{Bergstrom:1988fp,Rudaz:1989ij,Bouquet:1989sr,Buchmuller:2007ui}.  Forthcoming satellite \cite{CALET,DAMPE_TANSUO,Cumani:2015ava,Zhang:2014qga} and Cherenkov telescopes \cite{Abeysekara:2014ffg,CTA}, as well as the current Fermi Large Area Telescope (LAT) \cite{Ackermann:2013uma}, H.E.S.S.\ \cite{Abramowski:2013ax}, MAGIC \cite{Aleksic:2013xea} and VERITAS \cite{Zitzer:2015uta} instruments, will allow to probe this possibility with further sensitivity and towards higher energies. A similar `smoking gun' evidence of DM particles would be the observation of monochromatic cosmic neutrinos. For various reasons, this possibility has in general been less considered than the one of a gamma-ray line. 
We will here investigate the latest IceCube data available to search and constrain neutrino line (and line-like) signals from decaying DM.

On the theoretical side, one could in fact expect many decaying DM particle setups to give production rates of monochromatic neutrinos that are similar or larger than those into photon lines (see, e.g., \cite{Takayama:2000uz,Covi:2008jy,Chen:2008yi,Hisano:2008ah,Guo:2010vy,Garny:2010eg,Feldstein:2013kka,Higaki:2014dwa,Rott:2014kfa,Fong:2014bsa,Dudas:2014bca} and for counter cases, e.g., \cite{Kawasaki:1997ah,Arina:2009uq,Garny:2010eg,Gustafsson:2013gca,Ibarra:2013eda}).

On the experimental side, however, the traditionally reported energy resolution for cosmic neutrinos has been far from reaching the level of precision that holds for gamma-ray experiments. Still, this argument is not much relevant anymore as the neutrino energy resolution now reported by the IceCube collaboration is of the order of 15\,\% for cascade events \cite{Aartsen:2013vja}. It is thus close to what is achieved by current gamma-ray experiments in the GeV to TeV energy range.  Simultaneously, IceCube's effective area has increased, which leads to further improved sensitivity to neutrino fluxes, even after strong data cuts are applied to reduce atmospheric backgrounds. Still at the experimental level,  one should add that bounds usually quoted on the annihilation cross section and decay width are much less stringent for neutrino- than for gamma-lines. For example, in the case of a decay, the most stringent IceCube analysis reported so far \cite{Abbasi:2011eq} quotes a lifetime sensitivity of order ~$10^{25}$~sec at TeV energy, or few~$10^{26}$~sec above $10$~TeV, to be compared with the H.E.S.S.\ photon sensitivity of few~$10^{28}$~seconds within the $1-50$ TeV energy range \cite{Abramowski:2013ax,Gustafsson:2013gca}. Given the fact that no positive signal has been reported for a gamma-line at these energies, such a sensitivity comparison would imply that only models which predict much more monochromatic neutrinos than photons could be probed by neutrino detectors. The main purpose of this work is to show that this last statement does not hold anymore for decaying DM.

In our dedicated line search, we will include energy dispersions and optimize the statistical method for analyzing the energy spectrum. From recently released IceCube data \cite{Aartsen:2014muf,Icecube2015}, we then determine bounds on the flux of monochromatic neutrinos that could have been emitted by DM decay in the 1~TeV to 100~PeV energy range.  Above 100~TeV DM masses, our bounds are similar to the ones recently derived in Refs.~\cite{Rott:2014kfa,Esmaili:2014rma} using IceCube's  high-energy dataset from \cite{Aartsen:2014gkd}. Below these energies, and above few TeV, the bounds we obtain improve previous ones by more than one order of magnitude. As a result, there exists now a region of DM particle masses, in the range from several TeV to $50$~TeV, where stringent constraints on both monochromatic neutrinos and gammas do exist and are comparable. Interestingly, unlike at higher energies, this region is relevant for thermal DM candidates.

\section{Bounds on monochromatic neutrinos}

The IceCube collaboration recently released a data set of high-energy neutrinos collected during 2 years, from 2010 to 2012 \cite{Aartsen:2014muf}. At the highest energies, the neutrino events partly overlap with a longer exposure -- 3 years -- data set published few months earlier \cite{Aartsen:2014gkd}. The new data set has nevertheless the advantage that it extends down to lower energies, which can be of great interest for DM searches. From public IceCube data, it is therefore now possible to derive the  strongest current bounds on neutrino signals from DM decays in the TeV to PeV DM particle mass range.

In this study, we will take advantage of the latest data of Ref.~\cite{Aartsen:2014muf} and carefully derive upper bounds on a monochromatic neutrino line signal from decaying DM. The bounds we obtain will be compared to the current corresponding limits published by the IceCube collaboration  \cite{Abbasi:2011eq} below 100~TeV  and to those in Refs.~\cite{Rott:2014kfa,Esmaili:2014rma}, which instead used the 3-year high-energy data \cite{Aartsen:2014gkd}, above 30 TeV energies. Other previous bounds can be found in Refs.~\cite{PalomaresRuiz:2007ry, Covi:2009xn, Hisano:2008ah, Grefe:2011dp, Esmaili:2012us,Murase:2012xs,Bhattacharya:2014vwa}. 

\begin{figure}[t]
\center{\includegraphics[width=0.98 \columnwidth]{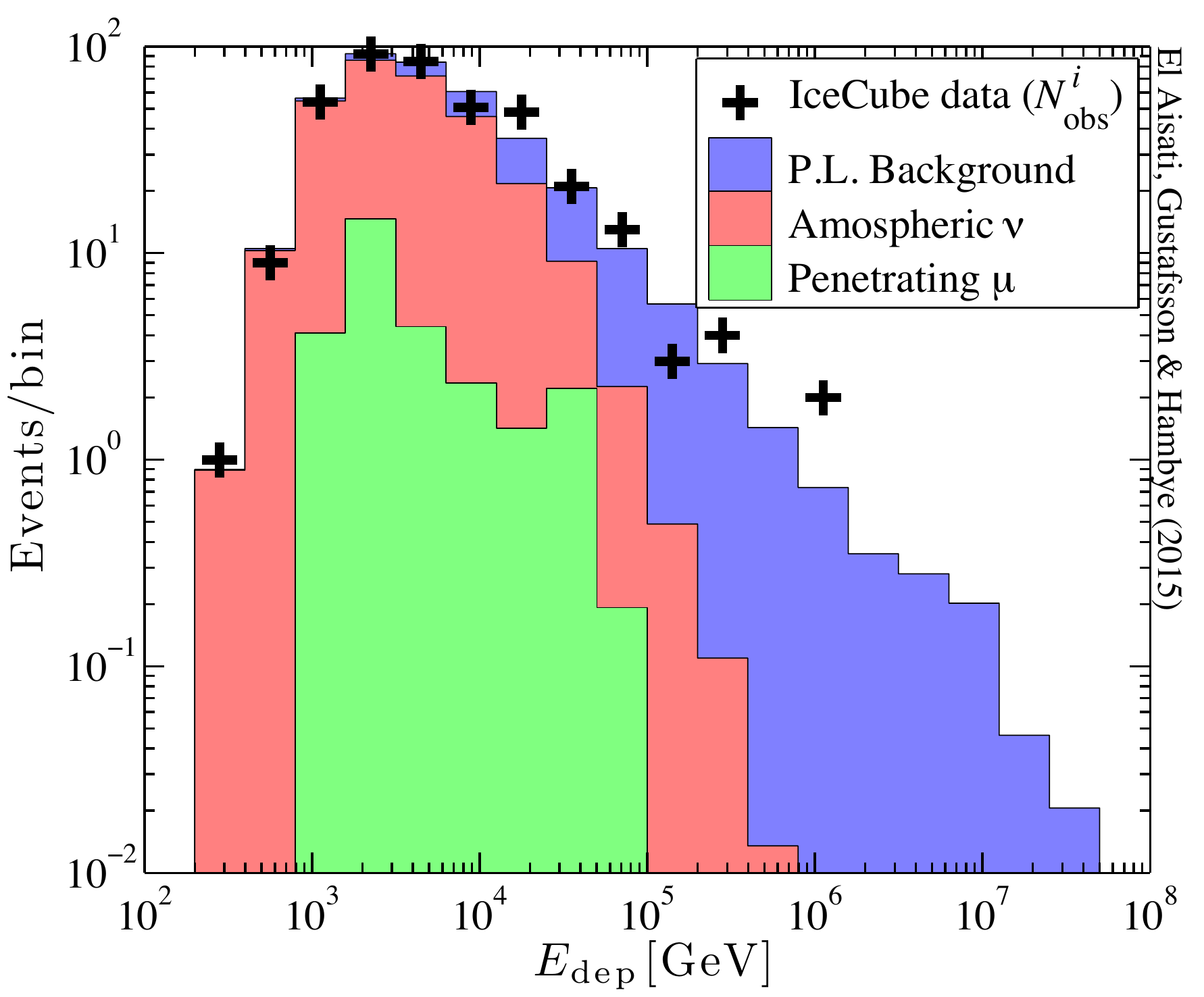}}
\caption{Deposited energy spectrum from the full sky (black curve)  as measured by IceCube. 
The colored regions show the expected atmospheric muon (bottom green) and neutrino (middle red) fluxes as well as the best-fit astrophysical neutrino power-law background model contribution (top blue). Data taken from Ref.~\cite{Aartsen:2014muf}.}  
\label{fig:data}
\end{figure}

The energy range covered by the data release in Ref.~\cite{Aartsen:2014muf} goes from 100 GeV to $10^8$~GeV. Twenty energy bins have been considered within this range with in total $N_{\rm tot} =383$ detected events. The number of events in each bin $i$ is denoted by $N_\mathrm{obs}^i$, and is shown in Fig.~\ref{fig:data}.
We will derive limits on DM signal by combining this neutrino data with the information on the instrument response  given in the supplementary material \cite{Icecube2015} to Ref.~\cite{Aartsen:2014muf}. The search for a possible detection of a monochromatic neutrino line and an improved approach to derive limits will be done in Section~\ref{sec:TS}.

\subsection{Dark matter signal prediction}

If the DM particle decays into neutrinos, with associated lifetime $\tau_{DM}$, it gives rise to a neutrino signal both from the Milky Way DM halo and from extragalactic DM. The former contributes with an intensity per solid angle $\Omega$ and neutrino energy $E_\nu$ that is
\be
\frac{{\rm d}\phi_{\mathrm{h}\alpha}}{{\rm d}E_\nu {\rm d}\Omega} (b,l)= 
  \frac{1}{4\pi\,m_{\rm DM}\,\tau_{\rm DM}} 
  \frac{{\rm d}N}{{\rm d}E_\nu}
  \int_{\rm l.o.s.} \!\!\!\! {\rm d}s\; 
  \rho_{\rm h}[r(s,\psi{[b,l]})].
  \label{eq:halo}
\ee
The integration is performed along the line of sight, where $s$ represents the distance to the observer. This intensity depends on the sky direction, which  is represented by the dependence on Galactic coordinates $b$ and $l$. However, in the case of a spherically symmetric DM halo, the signal intensity only depends on the relative angle $\psi$ between a given sky direction and the halo centre direction. We also take the DM halo to coincide with the Galactic centre (GC) at $(b,l)=  (0,0)$. The distance from the GC is given by $r(s,\psi) = \left( r_\odot^2+s^2-2 r s \cos\psi \right)^{1/2}$ and we take the Sun to GC distance to be $r_\odot =8.33$~kpc \cite{Gillessen:2008qv,Malkin:2012nu}.
For the DM density we use a NFW profile \cite{Navarro:1996gj},
\be
\rho_h(r) = \frac{r_\odot}{r} \left(\frac{a+r_\odot}{a+r}\right)^2 \rho_\odot,
\ee
with $\rho_\odot = 0.39$ GeV/cm$^3$ and $a=24$~kpc \cite{Catena:2009mf,Cirelli:2010xx,Read:2014qva}. ${\rm d}N/{\rm d}E_\nu$ is the neutrino-energy spectrum at the source, which in the case of  one monochromatic neutrino per DM particle decay would be ${\rm d}N/{\rm d}E = \delta(E-E_{\rm line})$.\footnote{\label{note1}
Doppler shift and electroweak correction effects broaden monochromatic lines. Typically, this broadening is significantly less than telescopes' energy resolutions. Though, for neutrino energies orders of magnitude larger than the $W/Z$ boson mass, electroweak corrections can be sizable.  At leading order, for example, $Z$-strahlung broadens a line so that only 99(89)\,\% of its original intensity remains within a 15\,\% energy band around its peak for $m_\text{DM} = 10^3 (10^8)$ GeV [using $D_{{\nu}\rightarrow \nu}$ in Eq.~(11) of \cite{Ciafaloni:2010ti} for the neutrino line shape, with $m_V\!=\!m_Z$ and $z\!=\!2 E_\nu/m_\text{DM}$ in their equation].  For large DM masses, though, higher-order resummation techniques might be required \cite{Berezinsky:2002hq,Kachelriess:2009zy}. We will not include these electroweak corrections.}

As for the extragalactic DM signal, it has an isotropic intensity given by
\be
\frac{{\rm d}\phi_\mathrm{eg}}{{\rm d}E_\nu d\Omega} \!=\! 
    \frac{\Omega_{\rm DM}\rho_{\rm c}}{4\pi m_{\rm DM} \tau_{\rm DM}}
    \int_0^\infty \!\!\!\!{\rm d}z\,
    \frac{c}{H(z)}
   \;\frac{{\rm d}N}{{\rm d}E} \Big|_{E=E_\nu(1+z)},
     \label{eq:eg}
\ee
where $H(z)=H_0 \sqrt{\Omega_\Lambda+\Omega_{\rm m}(1+z)^3}$ is the Hubble expansion rate as a function of redshift $z$, $c$ is the speed of light in vacuum and $\rho_{\rm c}=4.769\times10^{-6}\,{\rm GeV}\, {\rm cm}^{-3}$ denotes the critical density of the Universe. We furthermore assume a $\Lambda$CDM cosmology with parameters $\Omega_\Lambda=0.6844$, $\Omega_{\rm m}=0.3156$, $\Omega_{\rm DM}=0.2647$ and $h\equiv H_0/100\,{\rm km}\,{\rm s}^{-1}\,{\rm Mpc}^{-1}=0.6727$ --- as determined from the {\sl Planck} satellite's observations and found in Table~3 of Ref.~\cite{Ade:2015zuv}. An estimate of cosmological neutrino opacity reveals that emission below redshift $z \lesssim 10^4 (E_\nu/{\rm TeV})^{-2/7}$ undergoes negligible attenuation  \cite{Gondolo:1991rn}. We therefore neglect absorption effects in Eq.~(\ref{eq:eg}).

Eq. \eqref{eq:eg} shows that even if DM particles only decay into monochromatic neutrinos at the source, the {\it observed} cosmological signal ${d \phi_\mathrm{eg}}/{dE_\nu d\Omega}$ is \textit{not} monochromatic due to the redshifting effects. Instead of a monochromatic line at $E_{\rm line}$, it gives a broader line-like spectrum  peaking around $E_{\rm line}/2$. For $\psi \gtrsim 90^\circ$ the cosmological DM signal intensity is larger than that from the Galactic DM halo if integrated over an energy range from $E_{\rm line}/2$ to $E_{\rm line}$.  However, at lower latitudes the Galactic contribution dominates strongly and, e.g., at $\psi =10^\circ$ it is 5 times larger than the extragalactic one. As a result, the Galactic signal always dominates when considering the full sky flux. 
For IceCube observations, the Galactic domination is further enhanced because IceCube leans towards the GC and high-energy (typically $E_\nu \gtrsim 40$ TeV) up-going neutrinos undergo non-negligible absorption effects when going through Earth.

\subsection{IceCube's instrument response} 
\label{sec:IRF}

The differential number of expected events in the IceCube detector can be evaluated by convolving the neutrino intensities from Eqs.~(\ref{eq:halo}) and (\ref{eq:eg}) with the detector's instrument response:
\be
 \frac{{\rm d}N_\alpha}
 {{\rm d}E_\nu {\rm d}\Omega {\rm d}E' {\rm d}\cos\theta' {\rm d}\phi'} 
 = 
 \frac{{\rm d}(\phi_\mathrm{h}+\phi_\mathrm{eg})_\alpha}{{\rm d}E_\nu {\rm d}\Omega} 
 \mathcal{E}_\alpha
 D_{\mathrm{eff},\alpha},
\label{eq:dN}
\ee
where $E'$, $\theta'$ and $\phi'$ are events' reconstructed energy, Zenith angle and azimuth angle at IceCube, respectively. We here also introduce an index $\alpha \in \{e,\mu,\tau,\bar{e},\bar{\mu},\bar{\tau}\}$ to keep track of the flavor composition of  incoming neutrino fluxes at Earth's surface and if they are neutrinos ($\nu$) or anti-neutrinos ($\bar\nu$).

In general, Eq.~(\ref{eq:dN}) gives a spatial-dependent energy spectrum (induced by the superposition of the spatial varying intensity of the Galactic DM signal with the isotropic, but cosmologically redshifted, spectrum).

The exposure $\mathcal{E}_\alpha = A_{\mathrm{eff},\alpha} T$ is the product of the effective area $A_{\mathrm{eff},\alpha}(E_\nu,\theta,\phi)$ and the exposure time $T =\int {\rm d}t = 641$ days.\footnote{We approximate  $A_{\mathrm{eff}}$ to be independent on the azimuth angle  $\phi(b,l,t)$ and that the detector has no explicit time dependence.} The  Zenith direction $\hat{z}$ at IceCube is, in Galactic coordinates, at $b_{IC}\!=\!  -27.4^\circ$ and $l_{IC}\!=\! 303^\circ$, which allows to express the angle relative to the Zenith by $\theta(b,l) =  \arccos(\hat{z} \cdot \hat{s})$ for a given line of sight unit-vector $\hat{s}(b,l)$.
An incoming neutrino, with sky direction ($\theta,\phi$) and true energy $E_\nu$ at the {\it surface} of Earth, has an expected probability distribution, in reconstructed deposited energy $E'$  and reconstructed sky direction ($\theta',\phi'$), given by the dispersion function $D^\alpha_{\mathrm{eff}}(E',\theta',\phi';E_\nu,\theta(b,l,t),\phi(b,l,t))$.\footnote{$D^\alpha_\mathrm{eff}$ is commonly normalized such that its integration over  $E'$, $\cos\theta'$ and $\phi'$ gives 1.}

A bin integrated version of the differential instrument response functions $( A_\mathrm{eff} D_\mathrm{eff} )_\alpha$,  pre-integrated over all $\phi$ and $\phi'$, as well as over certain bin ranges in $\theta$, $\theta'$ $E$ and $E'$, is published in the supplemental material \cite{Icecube2015} to Ref.~\cite{Aartsen:2014muf}. This is the instrument response function we use in this study.\footnote{The effective areas provided in Ref.~\cite{Icecube2015} are summed over all interaction channels.}  It is worth noticing that there is a separate instrument response function for each neutrino flavor --- $e,\mu,\tau$ --- as well as if it is $\nu$ and $\bar{\nu}$, and that they show strong dependences on  IceCube's Zenith angle $\theta_{IC}$ due to energy losses and absorption effects in Earth for high energy neutrinos.

\medskip
For a given signal in Eq.~(\ref{eq:dN}), the expectation value of the differential number of observed neutrinos is
\be
  \frac{{\rm d}N^i_\alpha}{{\rm d}E_\nu d\Omega} 
  =  
  \int_{\Delta_i E'}        \!\!\!\!\!\!\!\!   {\rm d}E'       \!
  \int_{\Delta\theta'(t)}  \!\!\!\!\!\!\!\!\!\!  {\rm d}\cos\theta'   \!
  \int_{\Delta\phi'(t)}      \!\!\!\!\!\!\!\!\!\! {\rm d}\phi' 
   \;\frac{{\rm d}N_\alpha} {{\rm d}E_\nu {\rm d}\Omega {\rm d}E' {\rm d}\!\cos\theta' {\rm d}\phi'}, 
    \label{eq:dNi}
\ee
where we integrate $E'$ over each energy bin $\Delta_i E'$ using the same binning as in the publication \cite{Aartsen:2014muf}. 
Because Ref.~\cite{Aartsen:2014muf} provides no data on $\phi'(t)$ and only north/south separation in $\theta'$, we integrate over all $\phi'$ and $\theta'$ without optimization with respect to the DM signal's particular morphology (i.e.\ $\Delta \theta'(t)$ and $\Delta\phi'(t)$ include the full sky).

This leads to our final expression for the expected number of observed DM neutrino events in each energy bin~$i$: \be
N^i_{\rm DM}(m_{\rm DM},\tau_{\rm DM}) 
= 
 \int\! {\rm d}E 
 \int_{4\pi}\!\! {\rm d}{\Omega} 
\sum_{\substack{\alpha= e,\mu,\tau, \\ \quad \bar{e},\bar{\mu},\bar{\tau}}} P_\alpha
 \frac{{\rm d}N^i_{\alpha}}{{\rm d}E_\nu {\rm d}\Omega}, 
 \label{eq:NiDM}
\ee 
where the incoming neutrino flux's flavor composition at the surface of Earth is set by the probabilities $P_\alpha$.
\begin{figure}[t]
\center{\includegraphics[width=0.98 \columnwidth]{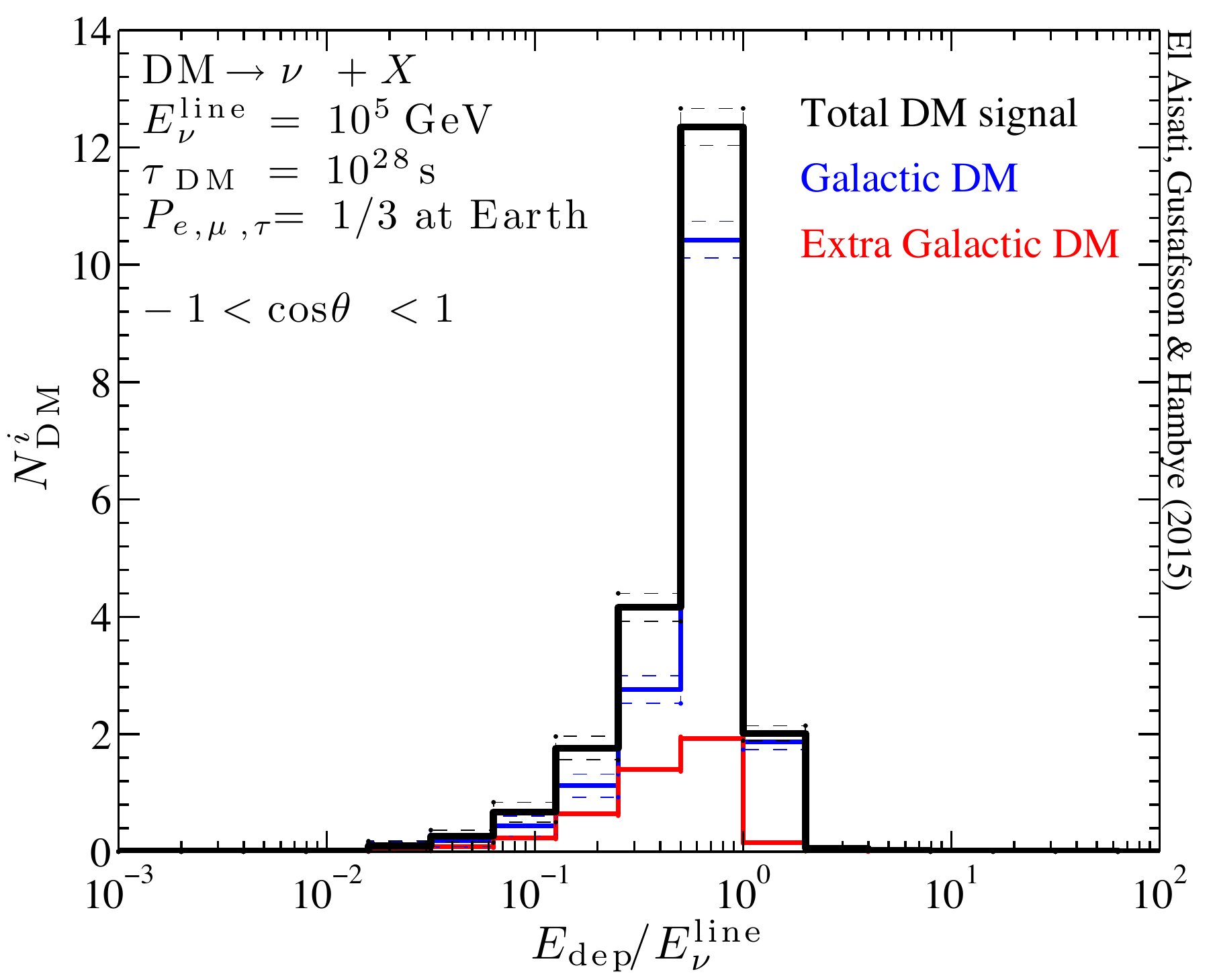}}
\caption{
Deposited-energy ($E_{\rm dep}$) spectrum for DM decay into monochromatic neutrinos of energy $E_\nu^{\rm line}=10^5$~GeV.
The signal normalization is for a DM particle with mass $2\times10^5$~GeV and lifetime $\tau_{\rm DM} = 10^{28}$s, with democratic neutrino flavor mixture and equal parts of $\nu$ and $\bar\nu$ in the flux reaching Earth surface. The diagram shows the total number of events  (black curve) after integration over the full sky and including propagation effects through Earth. The dominating Galactic DM contribution (blue) and the extragalactic DM contribution (red) are shown separately. The dashed curves enclose the effective area uncertainty range when the presented uncertainties of IceCube's instrument response function in the different sky directions are propagated and added in quadrature. The energy binning is the same as in the published instrument response function \cite{Aartsen:2014muf}.} 
\label{fig:Deff}
\end{figure}
As an example, Fig.~\ref{fig:Deff} shows the energy dispersed signal for $100$~TeV monochromatic neutrinos from DM particle decays (i.e.\ $DM\rightarrow \nu + X$, where $X$ is any state with negligible mass and with no signal contribution in IceCube). In this example, the flavor composition was taken to be $P_{e,\mu,\tau}\!=\!1/3$ with equal parts of $\nu$ and $\bar\nu$.

In general, after propagation, the average flavor composition at Earth's surface is set by the neutrino mixing probabilities (using expressions for long baseline oscillation $L\gg E_\nu/\Delta m_\nu^2$ \cite{Kayser:2005cd}):
\begin{eqnarray}
P(\nu_e\leftrightarrow \nu_e)        &=&0.573,\quad P(\nu_e\leftrightarrow \nu_\mu)=0.277\nonumber\\
P(\nu_e\leftrightarrow \nu_\tau)    &=&0.150,\quad P(\nu_\mu\leftrightarrow \nu_\mu)=0.348\nonumber\\
P(\nu_\mu\leftrightarrow \nu_\tau)&=&0.375,\quad P(\nu_\tau\leftrightarrow \nu_\tau)=0.475\nonumber.
\end{eqnarray} 
For the above results the following mixing angles have been used: $\sin^2 \theta_{12}=0.304$, $\sin^2 \theta_{23}=0.452$, $\sin^2 \theta_{13}=0.0218$ \cite{Gonzalez-Garcia:2014bfa}. Thus,  a pure flavor state $\beta$ produced at source gives $P_\alpha = P(\beta \leftrightarrow \alpha)$ in Eq.~(\ref{eq:NiDM}). Depending on the flavor of the neutrinos emitted, the bounds on the DM lifetime  vary by less than a factor two and for a pure $\nu$ or $\bar{\nu}$ flux the limits can vary a bit more --- we comment more on this in Appendix~\ref{sec:MoreLimits}.

\subsection{Robust dark matter constraints}
\label{sec:robust}

From the observed number of neutrino events in  each energy bin  $N_i$  (given in Fig.~\ref{fig:data}), robust upper bounds can be put on any model's neutrino flux prediction.

At the  $q=95$\,\% confidence level (CL), an upper bound $N^i_\mathrm{limit}$ in each bin is, by Neyman's construction \cite{Neyman:1937} from a Poisson distribution, given by\footnote{We note that our classical frequentist approach differs from the Bayesian approach in Ref.~\cite{Esmaili:2014rma} (and potentially also from \cite{Rott:2014kfa}), but numerically the results are practically identical except at 1-2 TeV DM masses where their Bayesian approach would give limits marginally weaker by 30\,\%.}
\be
\sum_{k=0}^{N^i_\mathrm{obs}} \frac{(N^i_\mathrm{limit})^k}{k!} e^{-N^i_\mathrm{limit}} = 1-q.
\ee

There is currently no certain understanding of the astrophysical neutrino background contribution at these energies. Therefore, we will only include the expected muon induced and atmospheric neutrino backgrounds. These background were derived in Ref.~\cite{Aartsen:2014muf}) and the sum of their central values $N_\mu$ and $N_\nu$, respectively, are represented by the upper edge of the red region in Fig.~\ref{fig:data}. To derive robust limits, we scale down these backgrounds to their 2-sigma lower edges, such that $N^i_{\rm bkg} = 0.538 N^i_\mu  + 0.938 N^i_\nu$. The 1-sigma uncertainties of the penetrating muon $N^i_\mu$ and atmospheric $N^i_\nu$ backgrounds are read from Table~1 in Ref.~\cite{Aartsen:2014muf}. The lower bound $\tau_\mathrm{limit}$ on the DM lifetime is then set so that the DM plus background, $N^i_{\rm DM}+N^i_{\rm bkg}$, expectation values do not overshoot the experimental limits $N^i_\mathrm{limit}$ in any energy bin. 
For each DM mass we thus set
\be
\tau_\mathrm{limit} 
= 
 \min
  \{
  \tau_{\rm{DM}} \in   \mathbb{R}^+ |  	  
  \forall i 
  :
  N^i_{\rm DM} +N^i_{\rm bkg}  < N^i_\mathrm{limit}
  \}.
\ee
Note that $N^i_{\rm bkg} \leq N^i_\mathrm{limit} $ is always fulfilled in this data set.\footnote{A violation of this inequality would be interpreted as an artifact of setting the background to a fixed assumed known value -- whereas it should  be associated with a certain uncertainty -- or as a large statistical downward fluctuation in the data. Statistically, one could  also worry that the limits are derived from the most constraining bin. However, in practice it is at most two $E'$ bins that are relevant and a trial factor of two would at most make a 95\,\% CL drop to a $1-(95\%)^2 \simeq 90\,\%$ CL limit.}

\begin{figure}[t]
\center{\includegraphics[width=0.96 \columnwidth]{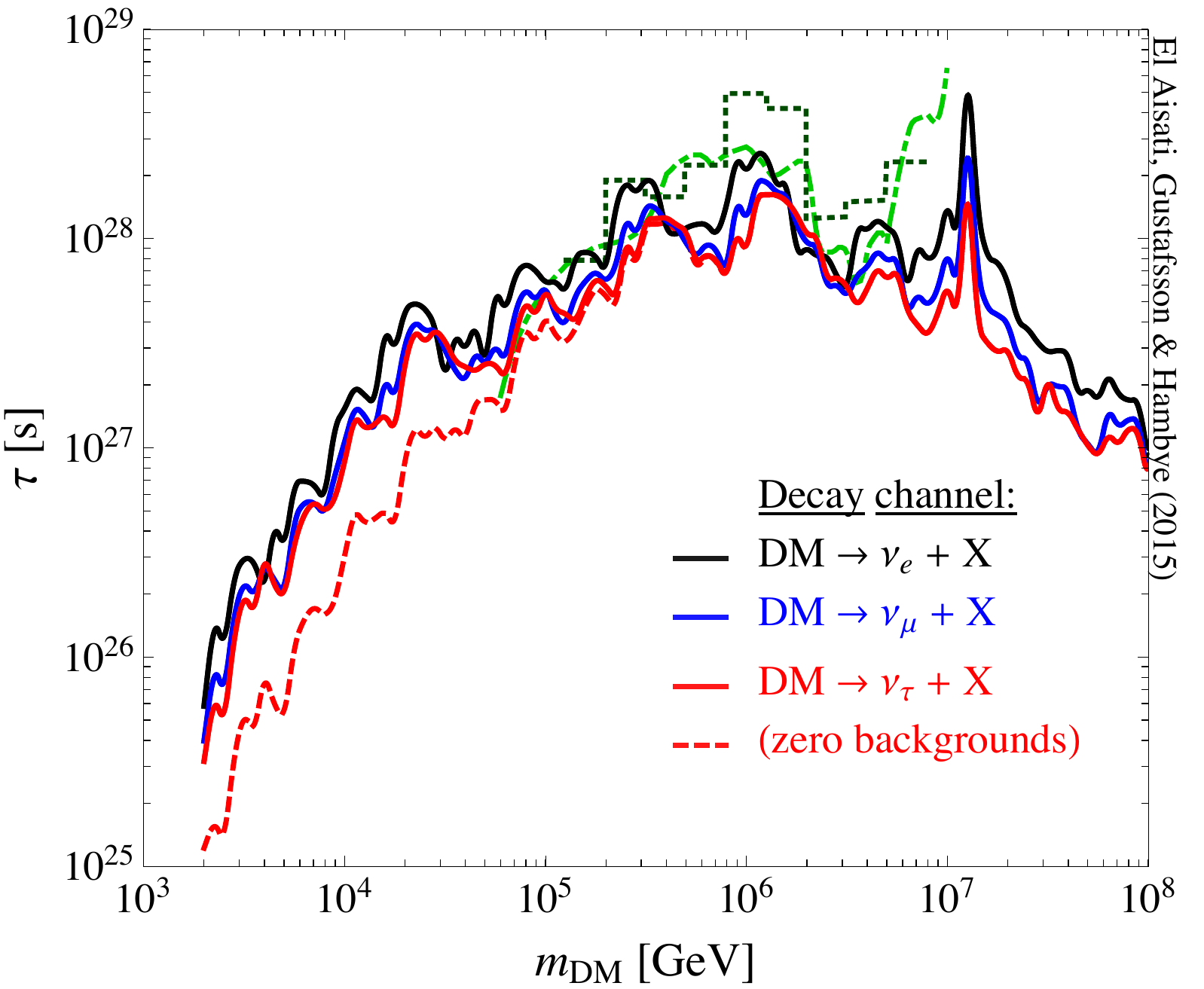}}
\caption{95\,\% CL limits on the lifetime of DM particle decay into monochromatic neutrinos. Solid (black, blue and red) curves are for DM decay into pure neutrino flavor states ($\nu_e$, $\nu_\mu$ and $\nu_\tau$, respectively) at production -- which after propagation to Earth are no longer pure flavor states.  The companion final state $X$ in the DM decay $DM\rightarrow \nu_\alpha + X$ is assumed to be light compared to the DM particle mass $m_{\rm DM}$. The dashed red curve shows the limit for ${\rm DM}\rightarrow \nu_\tau +X$ if no atmospheric background subtraction is made.  For comparison, we show the limits by Rott, Kohri and Park \cite{Rott:2014kfa} (90\,\% CL, green dotted curve) as well as Esmaili, Kang and Serpico \cite{Esmaili:2014rma} (90\,\% CL, green dashed-dotted curve) who analyzed the 3-years high-energy data set from Ref.~\cite{Aartsen:2014gkd}. Equal parts of $\nu$ and $\bar\nu$ are assumed for the DM signals.}
\label{fig:RobustLimit}
\end{figure}

In Fig.~\ref{fig:RobustLimit} we give 95\,\% CL bounds on DM decay into a monochromatic neutrino line.
We show  limits for the cases of pure $\nu_e$, $\nu_\mu$ or $\nu_\tau$ states and with equal parts of anti-neutrinos produced by the DM decay, ${\rm DM} \rightarrow \nu_{\alpha} + X$. For illustration, we also show the limits if zero background is assumed (thin dashed curve), although this is unrealistically conservative. 
In this figure, the bounds obtained are compared with the ones from Refs.~\cite{Rott:2014kfa} and \cite{Esmaili:2014rma}, in the range from 60 TeV to 10 PeV DM masses. For the DM signal, we assume equal induced fluxes of neutrinos and anti-neutrinos. The peak at $m_{DM} = 12.6$ PeV seen in the figure arises from the Glashow resonance of 6.3 PeV electron-antineutrinos annihilating on electrons in the ice \cite{Glashow:1960zz,Bhattacharya:2011qu}.

\subsection{Search for a $\nu$-line signal and improved bounds}
\label{sec:TS}

\subsubsection{Method}
\label{sec:Method}

Besides setting the best possible bounds on the DM lifetime, it is also of interest to search for any hint of a line signal in the data. 
We propose a novel methodology for neutrino line(-like) searches: a profile log-likelihood study of the energy spectrum including proper energy dispersions. This differs from previous studies that used Kolmogorov-Smirnov tests or approaches with no detailed spectral shape analysis \cite{Abbasi:2011eq,Abbasi:2012ws,PalomaresRuiz:2007ry, Covi:2009xn, Hisano:2008ah, Grefe:2011dp, Esmaili:2012us,Murase:2012xs,Bhattacharya:2014vwa,Rott:2014kfa,Esmaili:2014rma}. The method is similar to, e.g., the Fermi-LAT gamma-ray line search \cite{Ackermann:2013uma} (also briefly summarized in \cite{Gustafsson:2013fca}). This approach will enable us to search for line (and line-like) signals and further improve the DM bounds of the previous section.

To claim a detection of a monochromatic line requires enough statistics and energy resolution to distinguish such a feature in the spectrum from an assumed background model. A simple model to fit the neutrino flux that exceeds the expected muon and atmospheric backgrounds is an isotropic astrophysical signal with a single power-law energy spectrum, with equal parts of each flavor as well as of neutrinos and anti-neutrinos,
\be
\frac{{\rm d}\phi_{\mathrm{astro}}}{{\rm d}E_\nu {\rm d}\Omega} 
= 3 \times 10^{-18} \times \phi_0 \left( \frac{E_\nu}{E_0}\right)^{-\gamma} ,
\label{eq:astro}
\ee
where $E_0 = 10^5$ GeV. Such a power-law spectrum can be expected from conventional astrophysics contributions such as, e.g., active galactic nuclei and start burst galaxies with $2.0 \lesssim\! \gamma\! \lesssim 2.6$ \cite{Becker:2007sv,Loeb:2006tw}. The best-fit values from \cite{Aartsen:2014muf} are  $\gamma=2.46\pm0.12$  and $\phi_0=2.06^{+0.35}_{-0.26}$ GeV$^{-1}$cm$^{-2}$sr$^{-1}$s$^{-1}$.
To quantify the goodness-of-fit, we performed a Pearson $\chi^2$ test \cite{Cowan},
\be
 \chi^2 = \sum_{{\rm bins}\;\, i} \frac{(N^i_{\rm obs} - p_i N_{\rm tot})^2}{p_i N_{\rm tot}}.
 \label{eq:pearson}
\ee
Here, the set of $p_i = N^i_{\rm IC} /N_{\rm tot}$ gives the deposit-energy probability distribution of the model. The $N^i_{\rm IC}  = N^i_{\mu} + N^i_{\nu} + N^i_{\rm astro}$ are represented by the upper edge of the blue region in  Fig.~\ref{fig:data}. This will be our NULL model for generating Monte Carlo data. The IceCube data gives $\chi^2 = 12.7$ for this NULL model when we include all the 20 energy bins from Ref.~\cite{Aartsen:2014muf}. Pseudo experiments on $10^5$ Monte Carlo-generated data sets show that this corresponds to a P-value of 0.42.\footnote{A direct assumption of a $\chi^2$ distribution, with $N_{\rm bin} -5 = 15$ d.o.f., for the quantity in Eq.~(\ref{eq:pearson}) would instead have given a P-value of 0.63 for the NULL model. However, because of low statistics this assumption is not fully valid, and we generated data realizations from Poisson distributions of the number of events in each deposited-energy bin to calculate P-values. All fits to Monte Carlo data were done with an adjustable single power-law astrophysical component $N^i_\mathrm{\rm astro}(\phi_0,\gamma)$ together with free normalization of the two atmospheric background components $N^i_{\mu}$ and $N^i_{\nu}$ -- as in Eq.~\eqref{eq:model}. In Ref.~\cite{Aartsen:2014muf} the goodness-of-fit P-value is instead found  to be 0.2 for this model; but they consider three observables  (not only reconstructed deposited energy, but also directional and track property information) in their fits.\label{note:chi2}} 
The conclusion is that this models give a good fit and there is no apparent need for a contribution from a DM signal in the current IceCube spectrum.

\medskip
Given that such a simple background model fits the data well, we construct a test statistic (TS) by comparing the maximum log-likelihood of this type of background model to the hypothesis of a monochromatic neutrino-line signal on top of the background. Hence, to study if a line signal improves the fit, we evaluate
\be
 \text{TS}= 2 \ln \frac{ \mathcal{L}(n_{\rm sig} = n_{\rm sig,best} )}{ \mathcal{L}(n_{\rm sig} = 0)},
 \label{eq:TS}
\ee 
where the Poisson likelihood function is 
\be
\mathcal{L} = \prod_{{\rm bins}\; i } \frac{(N^i_\mathrm{model})^{N^i_{\rm obs}}}{N^i_{\rm obs} !} e^{-N^i_\mathrm{model}}
\ee
and the flux model $N^i_\mathrm{model}$ is the superposition of the, detector convolved, signal and backgrounds contributions 
\begin{align}
&N^i_\mathrm{model}(n_{\rm sig},n_{1,2,3},\phi_0, \gamma) = 
n_{\rm sig} N^i_\mathrm{DM}(m_{\rm DM},\tau_0) \nn\\
&\hspace{1.8cm}+ n_1 N^i_\mathrm{\mu}
+ n_2 N^i_\mathrm{\nu}
+ n_3 N^i_\mathrm{\rm astro}(\gamma,\phi_0).&
\label{eq:model}
\end{align}
The three backgrounds' normalizations $n_{1,2,3}$ and the astrophysical power-law spectral index $\gamma$ are set to their best-fit (largest likelihood) values for each value of a DM neutrino signal amplitude $n_{\rm sig}$. The normalizations of the DM $N^i_\mathrm{DM}$ and background $N^i_\mathrm{\rm astro}$ signals are relative to their (arbitrary) reference points of $\tau_0 = 10^{28}\;\text{s}$ in Eq.~(\ref{eq:NiDM}) and $\phi_0= 2.06$\;GeV$^{-1}$cm$^{-2}$sr$^{-1}$s$^{-1}$ in Eq.~(\ref{eq:astro}), respectively. We note that our best-fit (i.e.\ maximum likelihood) background-alone model has $n_1 = 1.36, n_2=0.85, n_3 =1.12$ and $\gamma=2.62$ in Eq.~\eqref{eq:model}. This background model is plotted in Fig.~\ref{fig:bestfit} of Appendix~\ref{sec:best} after convolution with the IceCube's instrument response.

\begin{figure}[t]
\center{\includegraphics[width=0.97 \columnwidth]{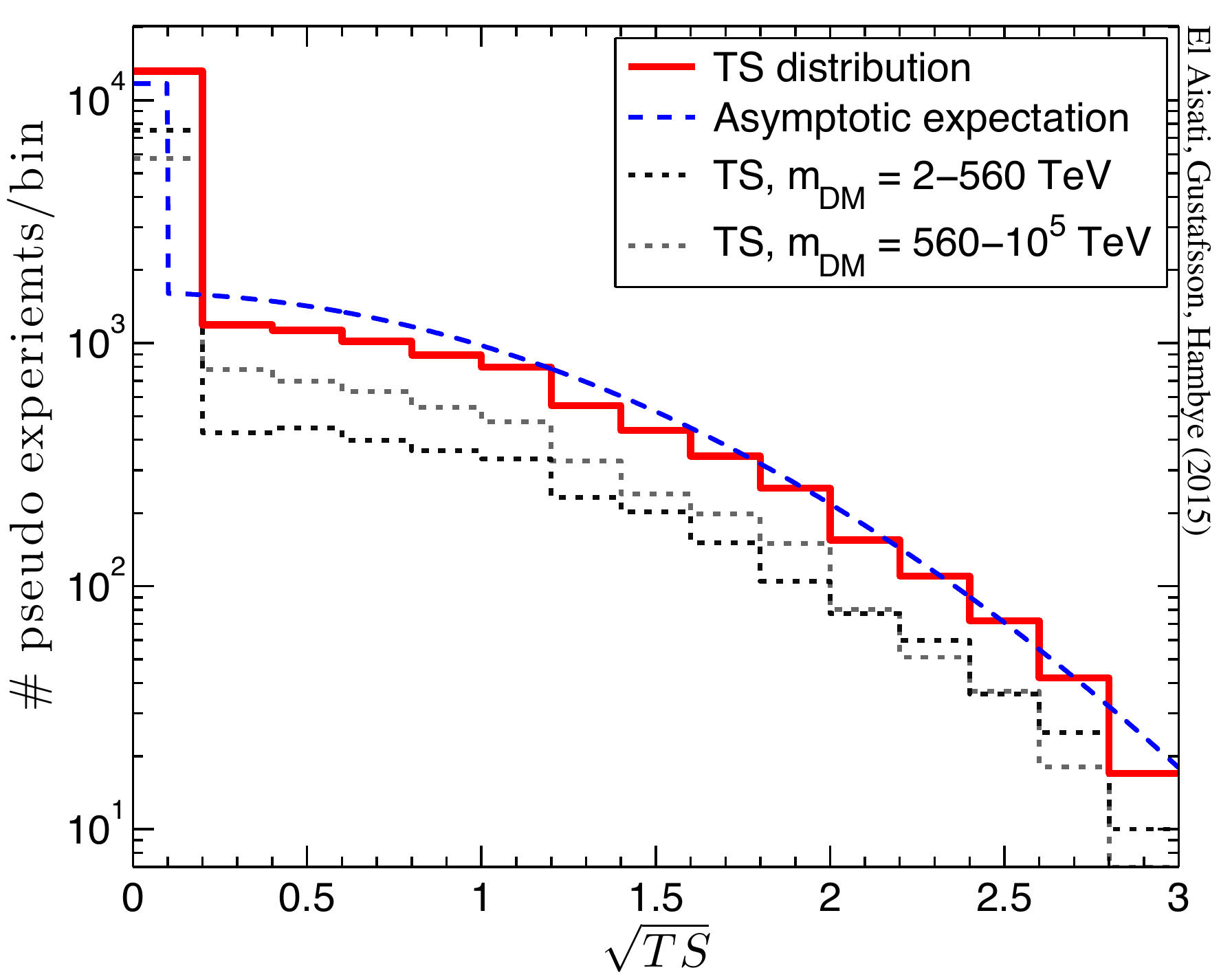}}
\caption{TS distribution from $2.02\!\times\!10^4$ monochromatic neutrino line searches (at our 101 used DM masses) in Monte Carlo generated pseudo data from our NULL background model assumption. The data follows well a $\frac{1}{2}\delta(\text{TS}) + \frac{1}{2} \chi^2(\text{TS})$ distribution (shown by the dashed blue line after multiplication by $2.02\times 10^4$). Dotted grey curves show the TS distribution from $m_{\rm DM}<560$~TeV (dark grey) and $m_{\rm DM}>560$~TeV (light grey) line searches separately.}
\label{fig:chi2}
\end{figure}

We use the full energy range of data for each TS determination. Because the statistics is fairly limited in all the 20 energy bins, we do not expect the systematic uncertainties to severely dominate over the large statistical uncertainties even if a single power-law model is assumed to be appropriate for the astrophysical background in this large energy range. By Wilks theorem \cite{Wilks1938} (or rather Chernoff theorem \cite{Chernoff1938}, as we constrain all signal contributors to have non-negative normalizations), the TS should asymptotically follow a $\frac{1}{2}\delta(\text{TS}) + \frac{1}{2} \chi^2(\text{TS})$  distribution.  However, due to the low statistics, this is not guaranteed and we performed $2.02\times10^4$ Monte Carlo pseudo experiments. For {\it each} DM mass range tested, we find good agreement between the obtained TS distribution  and the asymptotically expected distribution. Figure~\ref{fig:chi2} illustrates this by showing the collective TS distribution from our Monte Carlo data realizations of the background NULL hypothesis.

\subsubsection{Line search and limit results}

\begin{figure}[t]
\center{\includegraphics[width=0.97 \columnwidth]{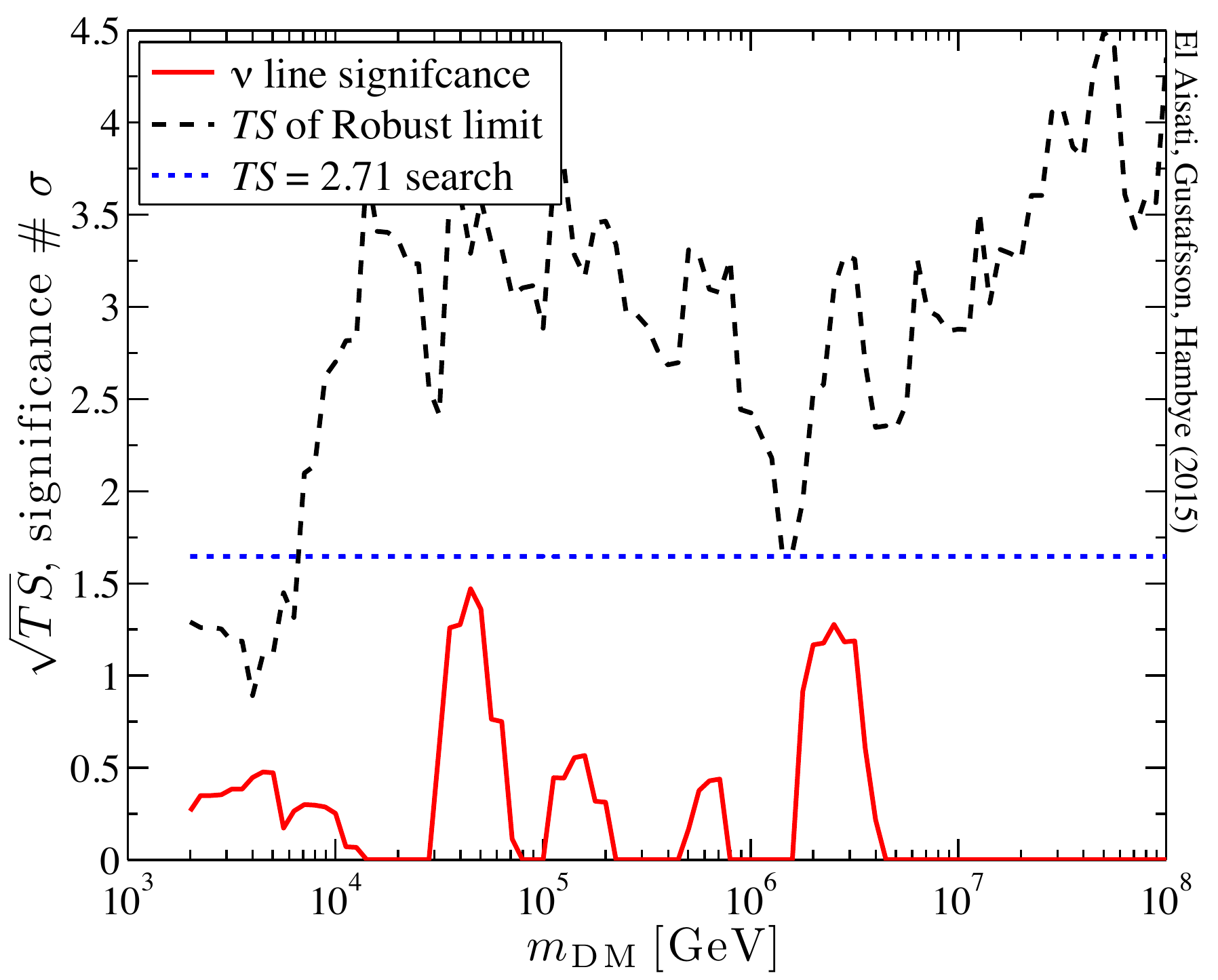}}
\caption{Local fit significance of a monochromatic line from DM decays {\it versus} DM mass (solid red curve). For comparison we also show the TS evaluated for signals at our `robust' limits with the method of Section~\ref{sec:robust} (black dashed) and our used profile Log-likelihood ratio limit at  $TS = 2.71$ (blue dotted curve).  Democratic flavor composition and equal parts of $\nu$ and $\bar\nu$ are assumed at Earth's surface.}
\label{fig:significance}
\end{figure}

No significant line signal was found in the IceCube data. We tested DM masses from 2 TeV to $100$ PeV in steps of 5\,\% in $\log_{10} E_\text{line}$ and checked the various flavors and neutrino/anti-neutrino compositions presented in the figures below and in Appendix~\ref{sec:MoreLimits}. The maximal TS was 2.9, corresponding to a Gaussian equivalence of  $\sqrt{\text{TS}} \simeq 1.7\;\sigma$ significance, and occurred for a line of pure $\nu_e$ at Earth and DM mass of 45 TeV. In Fig.~\ref{fig:significance}, we show a representative plot of the significances (solid red curve) as a function of DM mass. The plot is for the case of democratic flavor composition and equal parts of $\nu$ and $\bar{\nu}$ at the DM source. In this setup, the maximal significance is $\sqrt\text{TS} \simeq 1.5\;\sigma$ for $m_{\rm DM} = 44.8$~TeV when $\tau_{\rm DM} = 1.0\times10^{28}\;\text{s}$.
We note that the second largest TS peak, with $\sqrt\text{TS} \simeq1.3\,\sigma$ (for $m_{\rm DM} = 2.52$~PeV with $\tau_{\rm DM} = 3.9\times10^{28}\;\text{s}$), is due to the few neutrino events at $\sim1$\;PeV energies for which a DM origin has been speculated \cite{Feldstein:2013kka,Esmaili:2013gha,Ema:2013nda,Bai:2013nga,Bhattacharya:2014vwa,Higaki:2014dwa,Bhattacharya:2014yha,Ema:2014ufa,Rott:2014kfa,Esmaili:2014rma,Fong:2014bsa,Dudas:2014bca,Murase:2015gea}.

Our best-fit DM spectrum as well as our best $\sim$ 1 PeV line signal are shown for illustration in Fig.~\ref{fig:bestfit} of Appendix~\ref{sec:best}.
The 3-year IceCube data sample in \cite{Aartsen:2014gkd} and a DM signal with an accompanying continuum of lower energy neutrinos to a monochromatic neutrino line should not significantly alter these significances \cite{Fong:2014bsa}, unless the background modelling is very different.

\begin{figure}[t]
\center{\includegraphics[width=0.99 \columnwidth]{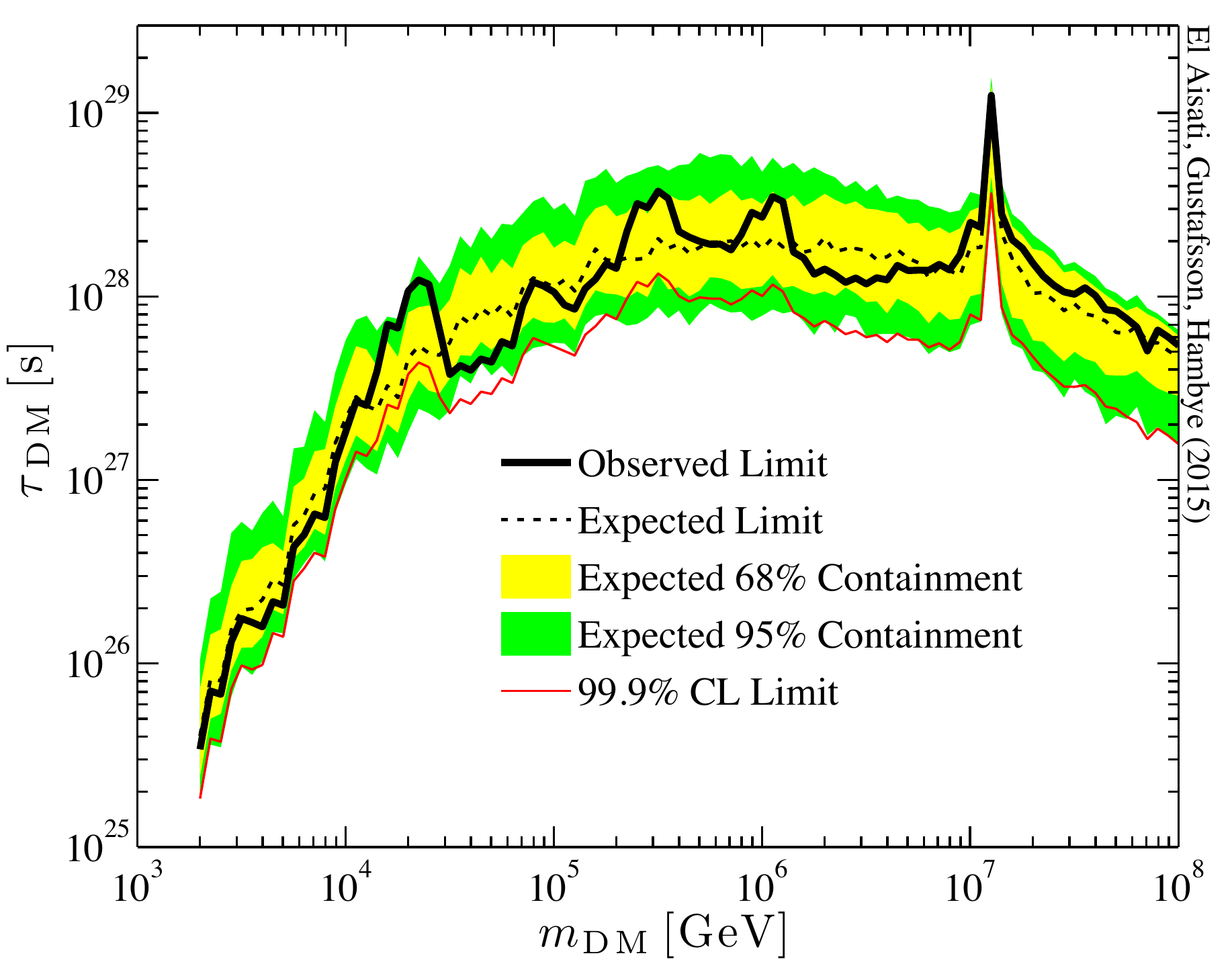}}
\caption{95\,\% CL lifetime limits (solid curve) on the DM particle decay lifetime into monochromatic neutrino. Expected sensitivity reach (dashed curve) and its 68\,\% (yellow) and 95\,\% (green) containment bands are also shown. The thin red line shows shows the 99.9\,CL limits. Democratic flavor composition and equal parts of $\nu$ and $\bar\nu$  are assumed for the incoming neutrino flux at Earth surface.}
\label{fig:ContainmentBand}
\end{figure}

With no significant line detected, we turn to derive stringent upper limits on DM decay into a monochromatic neutrino line. We set our 95\,\% CL lower limits on $\tau_{\rm DM}$, by requiring that the profile likelihood $\mathcal{L}(n_{\rm sig})$  with respect to its maximum values to have $\text{TS} < 2.71$ from Eq.~\eqref{eq:TS}.\footnote{ In Appendix~\ref{sec:coverage} we verified that this method has the statistically correct 95\,\%  coverage.}
The derived limits are shown in Fig.~\ref{fig:ContainmentBand}, where we also indicate by a thin red line the 99.9\,\% CL limit.
It turns out that this method improves lifetime constraints by up to a factor 5 compared to the approach of Section~\ref{sec:robust}. In fact, the shown 99.9\,\% CL limit follows fairly closely the `robust' limits from Fig.~\ref{fig:RobustLimit}.  All these limits are for the case of democratic neutrino flavor composition $P_\alpha = 1/3$ and equal part of $\nu$ and $\bar{\nu}$ (see however Appendix~\ref{sec:MoreLimits} for variations).

In Fig.~\ref{fig:ContainmentBand}, we also compare the derived limits with expected sensitivity. For data with low statistics, large fluctuations can naturally appear between different data realizations and  this comparison becomes particularly relevant. 200 Monte Carlo data sets are generated from the NULL hypothesis at each DM mass. The  expected 68\,\% (yellow band) and 95\,\% (green band)  limit containment bands are shown together with the median expectation (dashed black curve).

For illustration, we calculated the TS at the `robust' limit from Section~\ref{sec:robust} and show them in Fig.~\ref{fig:significance} by the dashed black curve (in this case we use democratic flavor composition of the DM signal, but otherwise the limit is as in Fig.~\ref{fig:RobustLimit}).
Except at the lowest masses, they always have TS values larger than 2.71 (which justifies to call them `robust'). The reason that they go below $\sqrt{\text{TS}} = \sqrt{2.71} = 1.65$ at the lowest DM masses is an effect from that our `robust' limit derivations assume a fixed atmospheric background contribution while the profile log-likelihood setup allows free normalization of all backgrounds. At the lowest line energies, it happened that our best-fits prefer lower atmospheric background contribution compared to the fixed NULL model from \cite{Aartsen:2014muf}.

\subsubsection{Constraints on other line-like signals}

The method described in Section~\ref{sec:Method} is fully general. In particular, many line-like spectra can pragmatically be parameterized as a single power-law with exponent $n$:
\be
\frac{{\rm d}N}{{\rm d}E} = \frac{2^{n+1} (n+1)}{m_{\rm DM}} \left(\frac{E}{m_{\rm DM}}\right)^{n}  \, \Theta\!\left({m_{\rm DM}}- 2E \right), 
 \label{eq:IB}
\ee
where the sharp spectral cut-off at $E\!=\!m_{\rm DM}/2$ is parameterized by the unit-step function \mbox{$\Theta\!\left(m_{\rm DM}-2E \right)$}.

This spectral shape works as a good proxy for several physical scenarios.  It appears in the cases of DM decay into three-body final states from effective operators (where $n\approx2\!-\!3$ \cite{ref:EffectiveNuLine})  or when DM particles decay into an intermediate state that then promptly decays into, e.g., neutrinos to form a `box-shaped' spectrum (where $n=0$ \cite{Ibarra:2012dw}). For {\it gamma-ray} DM signals, final state radiation and internal bremsstrahlung processes also give spectra approximately of this form (with $n\approx0-3$ \cite{Bergstrom:2004cy,Bergstrom:2005ss,Bringmann:2007nk}). Electroweak corrections to a monochromatic neutrino line, which we commented on in footnote~\ref{note1}, is however not of this form. Instead, the neutrino spectrum gets a low energy bump with a very steep rise towards its peak energy. Nonetheless, for all practical purposes this is equivalent to a monochromatic signal.

\begin{figure}[t]
\center{\includegraphics[width=0.99 \columnwidth]{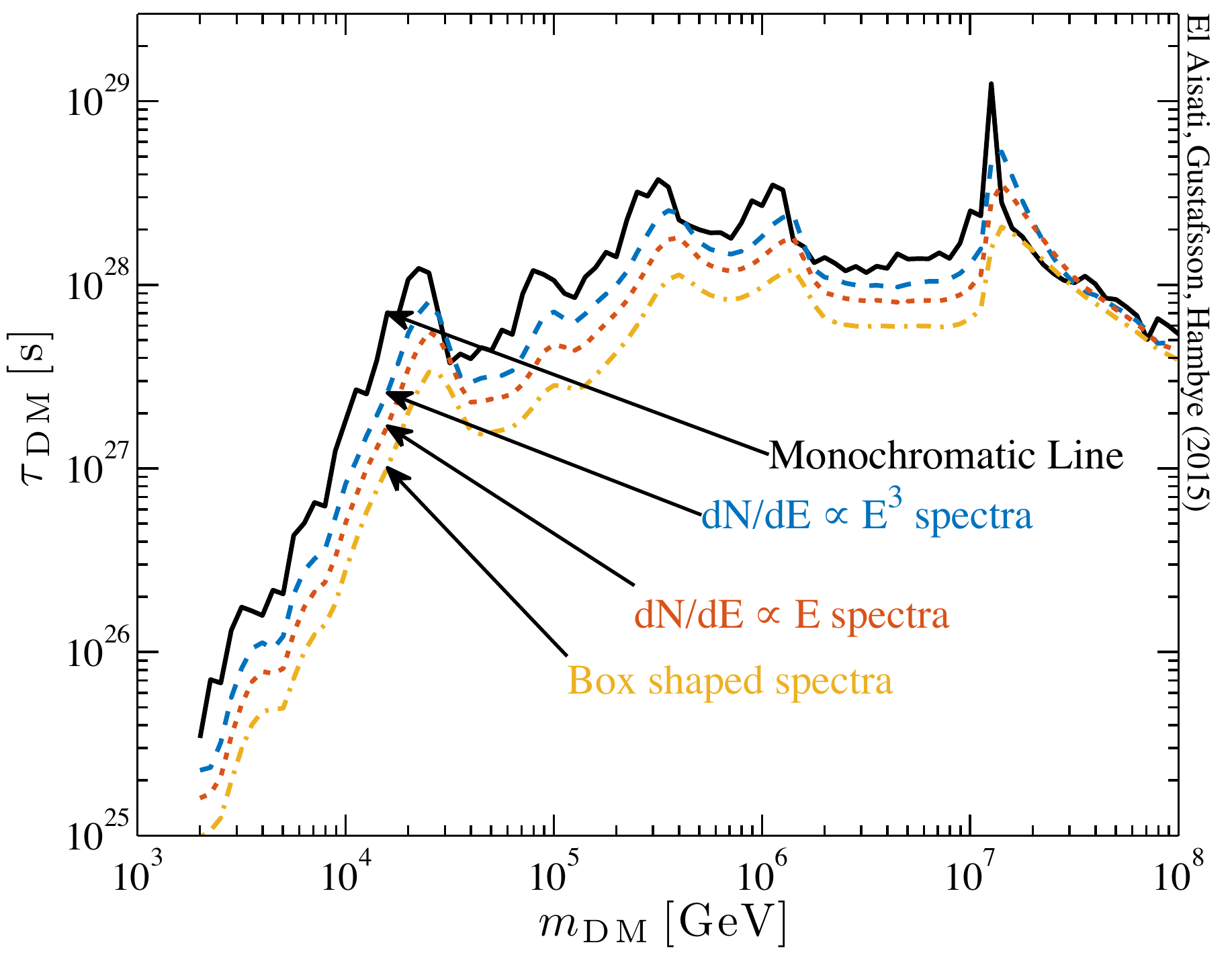}}
\caption{95\,\% CL lifetime limits on various neutrino line-like signals from DM decay: monochromatic line, internal Bremsstrahlung $dN/dE_\nu \propto E_\nu$ and $ \propto E_\nu^3$) and box-like spectrum ($dN/dE_\nu = \text{constant}$). Democratic flavor composition and equal parts of $\nu$ and $\bar\nu$ are assumed after propagation to Earth.}
\label{fig:IB}
\end{figure}

The spectra in  Eq.~\eqref{eq:IB} are normalized to give one neutrino per decaying DM particle. However, e.g., the box-shape spectrum is typically thought of as two intermediate states that promptly decay into two particles (in this case, two neutrinos), hence four neutrinos per DM particle decay with maximum energy of $m_{\rm DM}/4$ might be more natural in this case. 
We derive limits for the cases of $n=0,1,3$ for these types of DM signals and show those, together with the pure monochromatic line signal limit, in Fig.~\ref{fig:IB}.

\subsubsection{Possibilities of improvement} \label{sec:improvements}

In this study, we have presented a methodology to search and derive limits on a DM-induced neutrino line signal. We have always considered the sum of all events (e.g.\ track- plus cascade-like events), the full sky region and events binned in energy. Future improvement might be possible by using additional information in the data:
\begin{itemize}
\setlength\itemsep{0em}
\item `Morphology sorting' of the neutrino events, e.g, into cascade- and track-like events and correlate this to the expected DM and background signals \cite{Mandal:2009yk}. Typically, a muon neutrino produces a track-like signature, while tau- and electron-neutrino interactions have cascade-like morphologies \cite{Aartsen:2015ivb}. Utilizing the method of Section~\ref{sec:robust}, and assuming no backgrounds, limits from track and cascade events separately did not significantly improve limits.
\item Using directionality information of incoming neutrinos to, e.g., find an optimized sky regions where signal to background ratio is expected to be the largest \cite{Grefe:2011dp,Covi:2008jy}. Unfortunately, the publicly provided events and the effective areas have limited informations on this \cite{Icecube2015}. A simple check with separation into north and south hemisphere subsets did not reveal significantly improved limits. 
\item Exploring the data unbinned in reconstructed energies. This would require a better resolution of the instrument response function than the limited 20 energy bins provided between $10^2$ to $10^8$ GeV \cite{Icecube2015}. 
\end{itemize}
Improvements in sensitivity could therefore be possible, but care must be taken to not reduce the already low statistics in any analysis. Systematic effects, e.g. in the effective areas and the background modelling, might also be studied in more detail \cite{Abbasi:2011eq,Aartsen:2014gkd,Aartsen:2014muf}. A straightforward propagation of effective areas uncertainties, taken from \cite{Icecube2015}, by adding them in quadrature revealed less than 30\,\% rescaling of limits.

If we recall that this analysis is based on a data set collected only from year 2010 to 2012, the method approach presented here could even open up the potential to discover a DM induced neutrino line signal in already collected data with the used, or improved, event selection cuts.

\section{Conclusions}
\begin{figure}[t]
\vspace{-0.1cm}
\center{\includegraphics[width=0.96 \columnwidth]{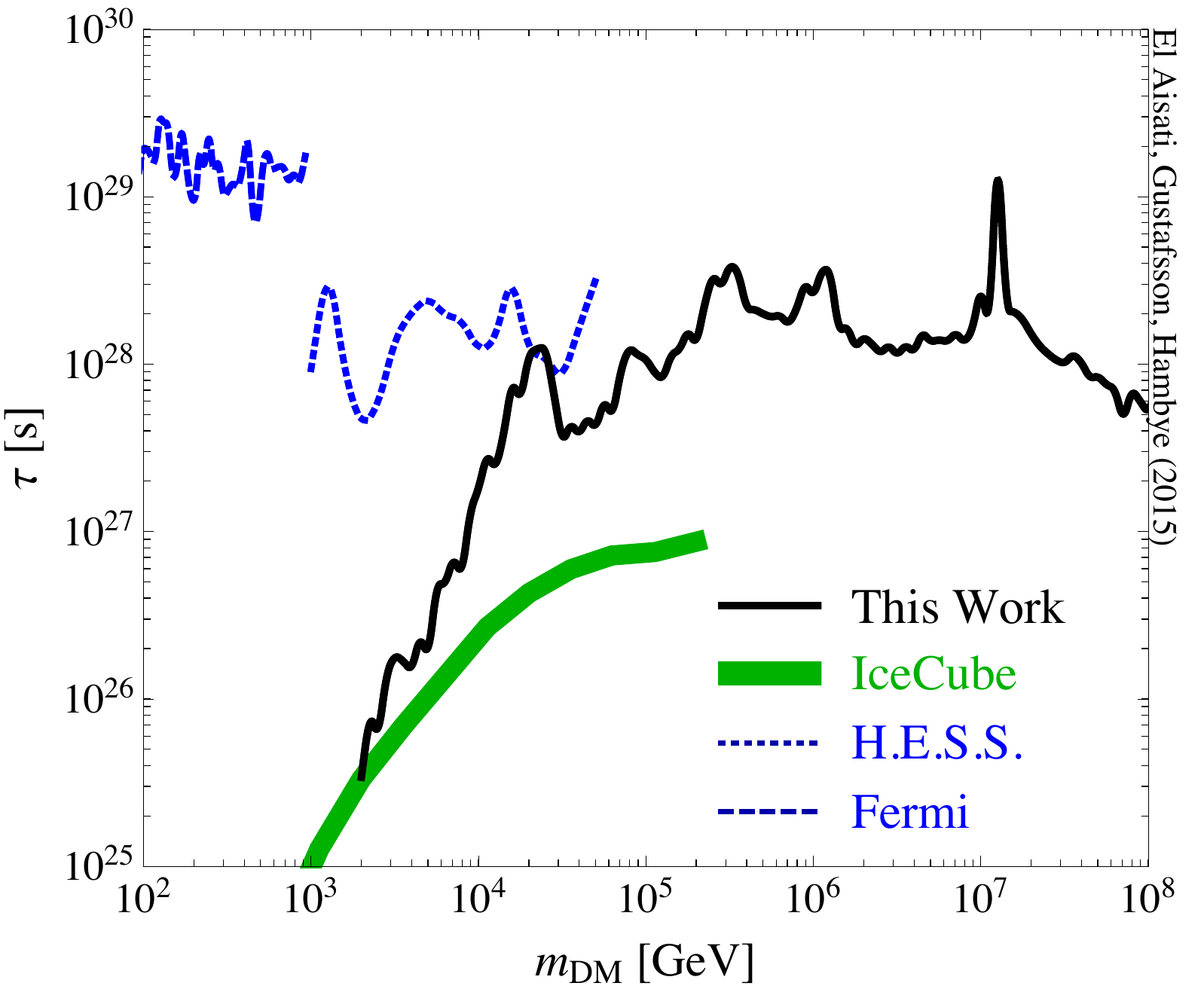}}
\caption{Lifetime limits on DM particle decay into monochromatic lines $\nu + \gamma$. From Fermi-LAT \cite{TheFermi-LAT:2015gja} (blue, dashed) and H.E.S.S. \cite{Abramowski:2013ax,Gustafsson:2013gca} (blue, dotted) using gamma-ray data compared to the neutrino line bounds derived in this study (solid black) as well as previous IceCube limits \cite{Abbasi:2011eq} 
(thick green).}
\label{fig:gammanubounds}
\end{figure}

We have presented in Figs.~\ref{fig:ContainmentBand} and \ref{fig:IB} new bounds on the lifetime of a DM particle decaying into monochromatic neutrinos or similar line-like signals. In Appendix~\ref{sec:MoreLimits}, we also show complementary limits, including various assumptions on the neutrino flux composition of the emitted signal. Above few TeV DM masses, our derived limits are significantly stronger than the latest limits from the IceCube collaboration \cite{Abbasi:2011eq}, which we illustrate in Fig.~\ref{fig:gammanubounds}.  For example, around 20 TeV DM masses our limits are a factor 25 better than the  IceCube collaboration's 90\,\% CL limits (that used a slightly different DM density profile) \cite{Abbasi:2011eq}.
For DM masses above 100 TeV and up to 1~PeV, our bounds are in agreement with those obtained in Refs.~\cite{Rott:2014kfa,Esmaili:2014rma} (which were compared to our `robust' limits in Fig.~\ref{fig:RobustLimit}). The improvements in limits are a consequence of  analyzing the latest data published by the IceCube collaboration \cite{Aartsen:2014muf, Icecube2015} and a refined approach to neutrino line searches by studying the energy spectrum with a profiled likelihood method.

For energies above a few TeV, the sensitivities reached on neutrino lines are now comparable to those existing on monochromatic gamma-ray lines. As illustrated in Fig.~\ref{fig:gammanubounds}, the following picture emerges:
\begin{enumerate}[i)]
\setlength\itemsep{0em}
\item For DM masses below few~TeV, constraints on decay lifetimes to gamma lines remain orders of magnitudes stronger than those to neutrino lines;
\item Above the maximum energy considered by the H.E.S.S.~experiment, $E_\gamma =25$~TeV, there are to our knowledge no numerically precise gamma-line constraints  (see however \cite{Atkins:2004jf, Abdo:2009ku, Borione:1997fy, Schatz:2003aw, Abeysekara:2014ffg, Murase:2015gea}) but strong neutrino-line constraints exist now up to energies several orders of magnitude higher; 
\item  In the multi-TeV to 50 TeV mass range, the lifetime constraints for these two monochromatic decay channels only differ by a factor of one up to an order of magnitude.
\end{enumerate}
With foreseen improvements in both neutrino \cite{Aartsen:2013mla,Aartsen:2014njl,Avrorin:2013uyc,Aartsen:2014oha,Margiotta:2014eaa} and gamma-ray  \cite{Bringmann:2012ez,Bergstrom:2012vd,Conrad:2015bsa,Ibarra:2015tya} data, this opens up increased chances to see a `double-barreled smoking gun' signal in the form of monochromatic neutrino line plus gamma-ray line from DM particles \cite{ref:EffectiveNuLine}.

\begin{acknowledgements}
We acknowledge stimulating and enlightening discussions with C.~De Clercq, F.~Halzen and especially J.~L\"unemann and Jakob van Santen for IceCube data related support. We thank L.~Covi, T. Scarn\`a,  S.\ Schumann and M.~Tytgat for useful discussions. T.H.\ thanks the IPhT-CEA-Saclay group for hospitality, where part of this work has been done. M.G.\ acknowledges partial support from the European Union FP7 ITN Invisibles (Marie Curie Actions, PITN-GA-2011-289442).
This work is supported by the FNRS-FRS, the FRIA, the IISN, an ULB-ARC and the Belgian Science Policy, IAP VI-11.
\end{acknowledgements}

\appendix

\section{Best-fit spectra}\label{sec:best}

To illustrate how some of the best-fit deposit energy spectra compare to the IceCube data, we show in Fig~\ref{fig:bestfit}: 
\begin{figure}[b]
\center{\includegraphics[width=0.99 \columnwidth]{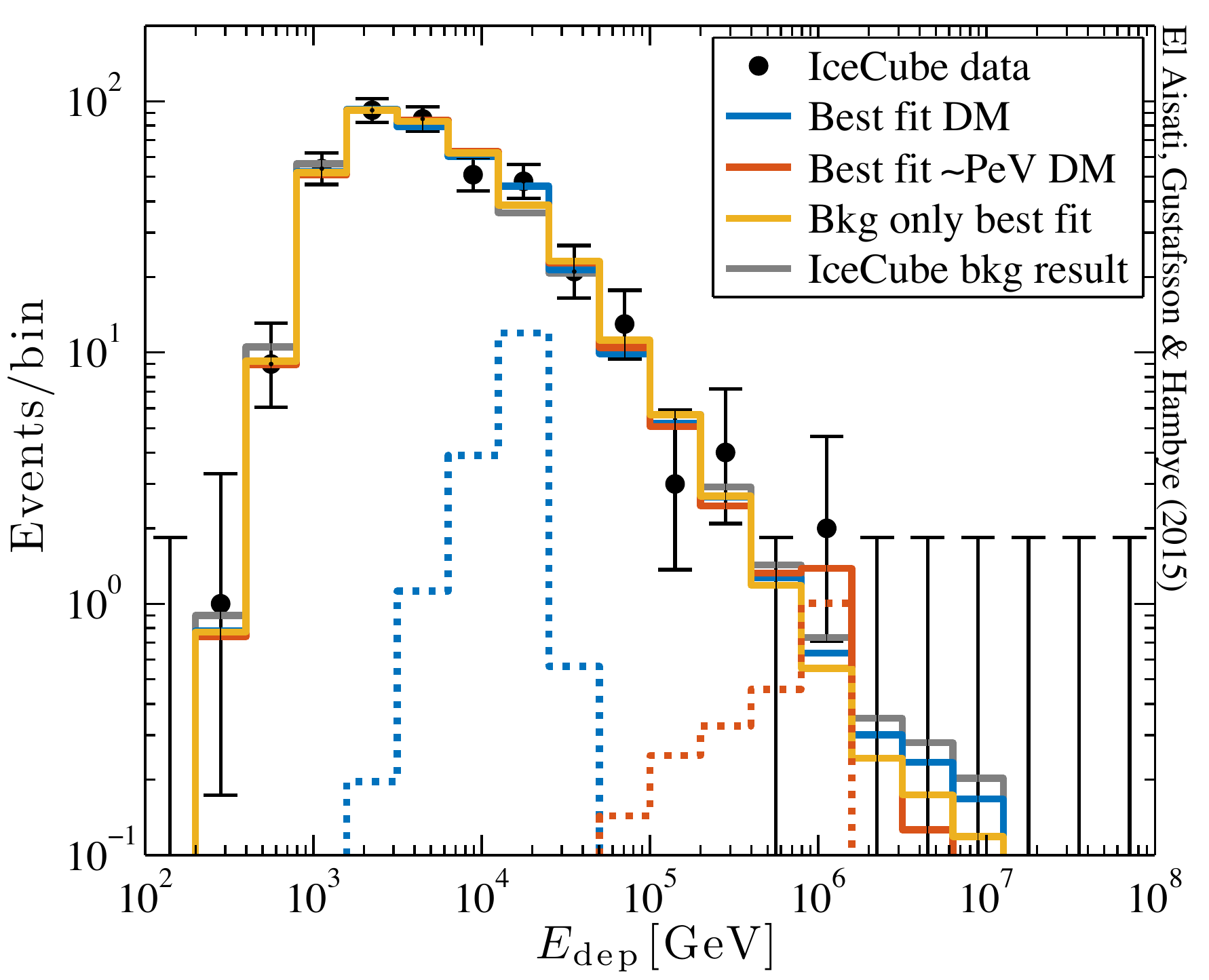}}
\caption{Observed deposit energy spectra compared to: our best-fit model with a DM particle mass of 44.8 TeV (blue, and blue-dotted for the DM contribution); our secondly best-fit  model with a DM particle mass of 2.52 PeV (red, and red-dotted for the DM contribution); our best-fit background-only model (orange); and the IceCube collaboration's \cite{Aartsen:2014muf} best-fit background model (grey). The best fit DM decay signal contributions are shown for the  44.8 TeV and  2.52 PeV DM particles by the  dashed red and dashed blue curves, respectively. DM models are for democratic flavor composition and with equal parts of $\nu$ and $\bar\nu$ flux at Earth surface. The significance for both the DM signals are less than $1.5\,\sigma$.}
\label{fig:bestfit}
\end{figure}
i) our best-fit background model [presented below Eq.~\eqref{eq:model}] ii) our best-fit DM signal model and iii) our best-fit DM model with a monochromatic neutrino line signal around the 2 observed events at $\sim 1$~PeV energy. These are spectra in the case of democratic flavor and equal parts of $\nu$ and $\bar\nu$.

The $\sim1$~PeV DM signal might seem too low to give the best-fit, but we checked that an increased signal would very slightly worsen the likelihood of the fit.  We also show the IceCube collaborations \cite{Aartsen:2014muf} best-fit model from Fig~\ref{fig:data}. Similarly to their best-fit background model (considered in Section~\ref{sec:Method}) \cite{Aartsen:2014muf}, our best fit background model gives a $\chi^2 = 12.4$ from Eq.~\eqref{eq:pearson}. This corresponds to a P-value of 0.40 from a Monte Carlo validation (as opposed to a P-value of 0.65 if a $\chi^2$ distribution were assumed -- c.f.\ footnote \ref{note:chi2}).

\section{Statistical coverage}\label{sec:coverage}

The statistical coverage of our limits was investigated by simulating a DM line signal on top of our NULL hypothesis (given in the beginning of Section~\ref{sec:Method}). We simulated $10^5$ Monte Carlo realizations,  and looked at 101 DM masses between 2 TeV and 200 PeV. The monochromatic neutrino lines were randomly given a signal strength corresponding to a $\tau_\text{DM}$ between one order of magnitude larger or smaller than our derived DM limits $\tau^\text{IC}_\text{limit}$. 

\begin{figure}[b]
\vspace{-1.7mm}
\center{\includegraphics[width=0.995 \columnwidth]{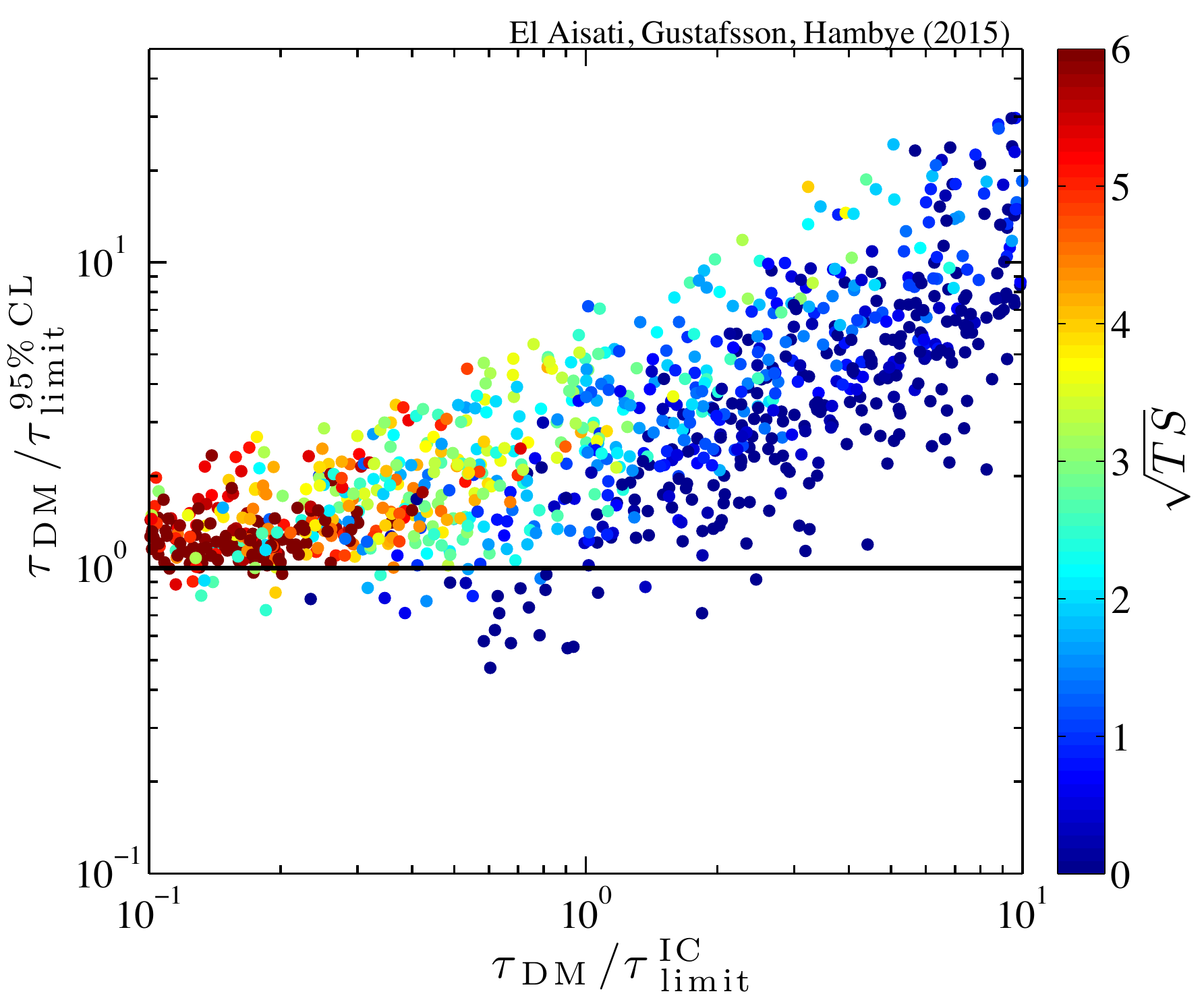}}
\caption{
Comparison of the 95\,\%\,CL lower limit $\tau_\text{limit}^\text{95\,\%\,CL}$ to the true DM lifetime value $\tau_\text{DM}$ for a set of 100 Monte Carlo realizations (for each 101 DM masses tested between 2\;TeV and 200 PeV). Each realization is from a model with a DM particle lifetime  $\tau_\text{DM}$ randomly drawn between 0.1 and 10 times the lifetime limit derived from the IceCube data $\tau^\text{IC}_\text{limit}$. The color of each point gives the found significance $\sqrt{\text{TS}}$ (i.e. the number of standard deviations $\sigma$) for the injected DM model. Points above the solid horizontal line represent realizations where the lower lifetime limit covers the true injected value. Democratic flavor and equal parts of $\nu$ and $\bar\nu$ composition was assumed at Earth surface for the DM signals.}
\label{fig:coverage}
\end{figure}

For lifetimes shorter than our derived limits from the IceCube data ($\tau_\text{DM}/\tau^\text{IC}_\text{limit} <1$), the found coverage is 93\,\%, which is in good agreement with our stated 95\,\% CL. For $\tau_\text{DM}/\tau^\text{IC}_\text{limit} >1$ the coverage is 99\,\%, which corresponds to a safe overcoverage that is expected for low signal strengths. In Fig.~\ref{fig:coverage} 
we show the first 1010 points, where we color-code each simulated DM signal with its corresponding TS value.  For a clearer color scale, we assigned points with $\sqrt{TS}>6$ the value 6.

\section{$\nu$  vs $\bar\nu$ and flavor composition dependences on DM lifetime limits}\label{sec:MoreLimits} 

In this appendix we collect some complementary limits that might be of interest. All limits in this Appendix are derived by the method described in Section~\ref{sec:Method}.

\begin{figure}
\center{\includegraphics[width=0.99\columnwidth]{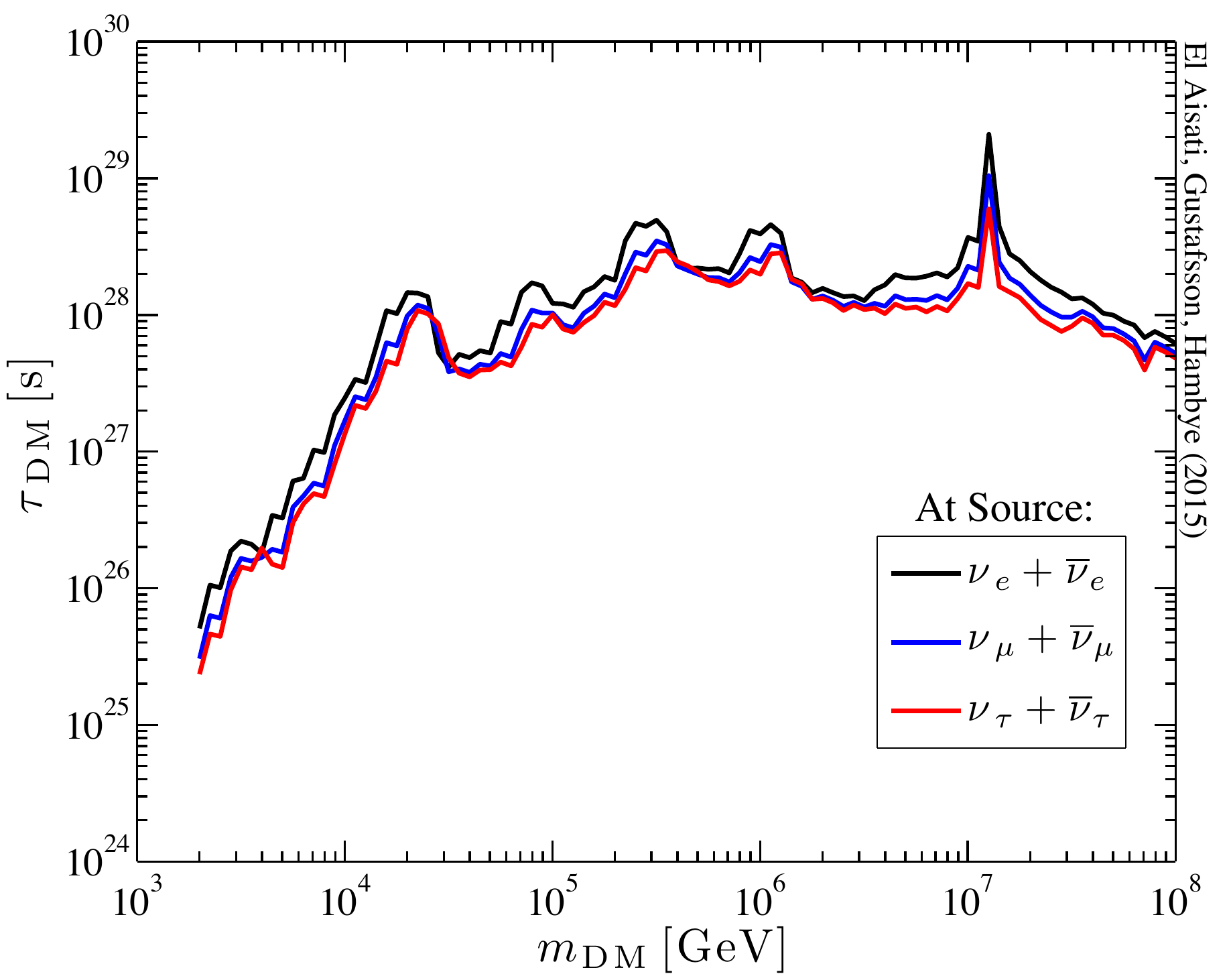}}
\caption{
95\,\% CL limits on the lifetime of DM particle decay into monochromatic neutrinos for pure $e$ (black), $\mu$ (blue) or $\tau$ (red) neutrinos injected at {\it Source} and with equal parts of $\nu$ and $\bar\nu$.}
\label{fig:Source}
\end{figure}

Figure~\ref{fig:Source} shows the 95\,\% CL lower limits on DM particle lifetime into a monochromatic neutrino in the cases of pure electron, muon or tau neutrinos injected at the DM source and with equal parts of $\nu$ and $\bar\nu$. The neutrino flavor oscillations from Section~\ref{sec:IRF} are used for the propagation from source to Earth's surface. 

\begin{figure}[t]
\center{\includegraphics[width=0.99 \columnwidth]{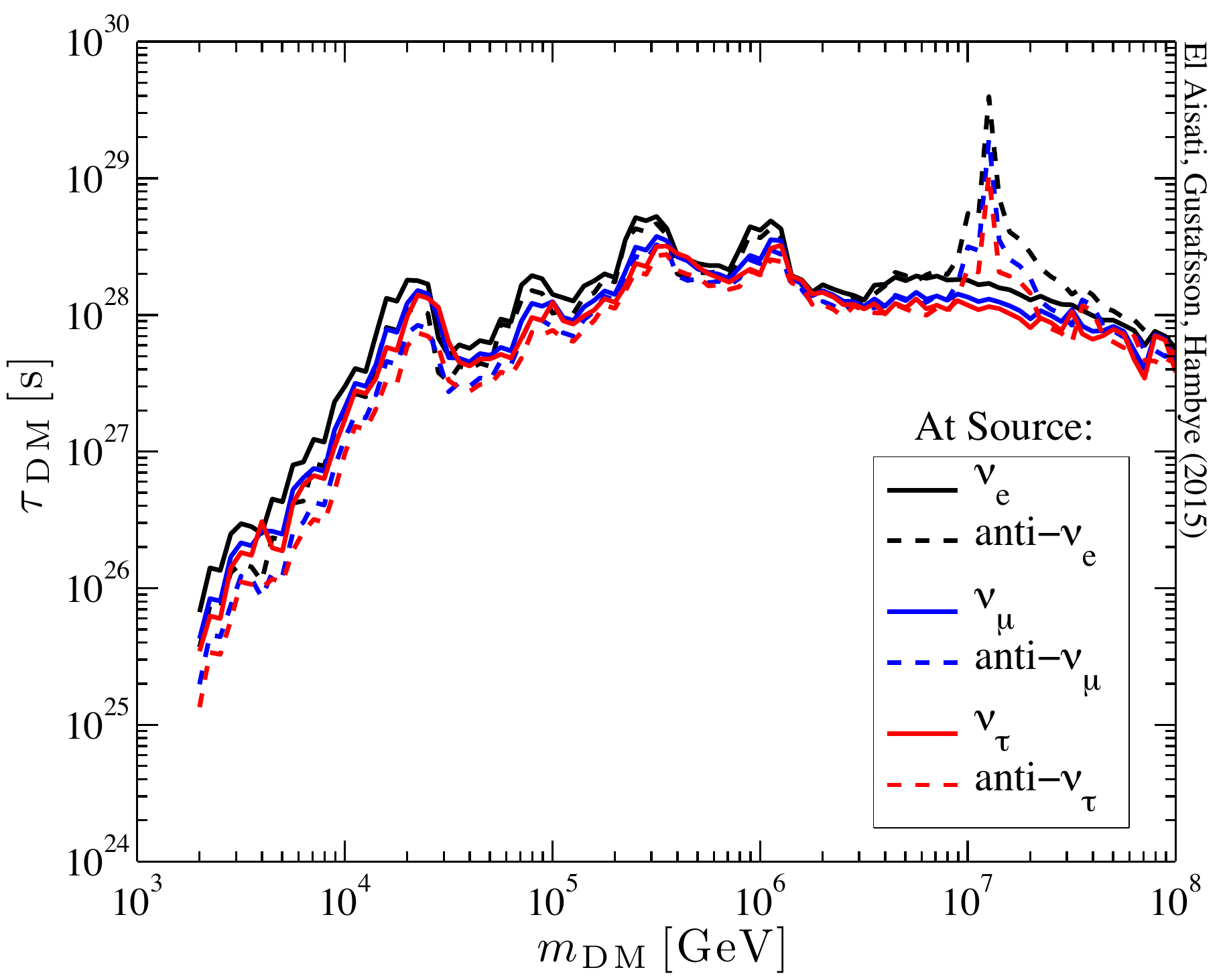}}
\caption{
Same as Fig.~\ref{fig:Source} with the pure $\nu_e$, $\nu_\mu$ and $\nu_\tau$ at {\it Source}, but also separated into neutrinos (solid) or anti-neutrinos (dashed)}
\label{fig:Source_anti}
\end{figure}

\bigbreak
In Fig.~\ref{fig:Source_anti}, we use the same setup, but assume only neutrinos or anti-neutrinos are injected by the DM decay. 

\begin{figure}
\center{\includegraphics[width=0.99\columnwidth]{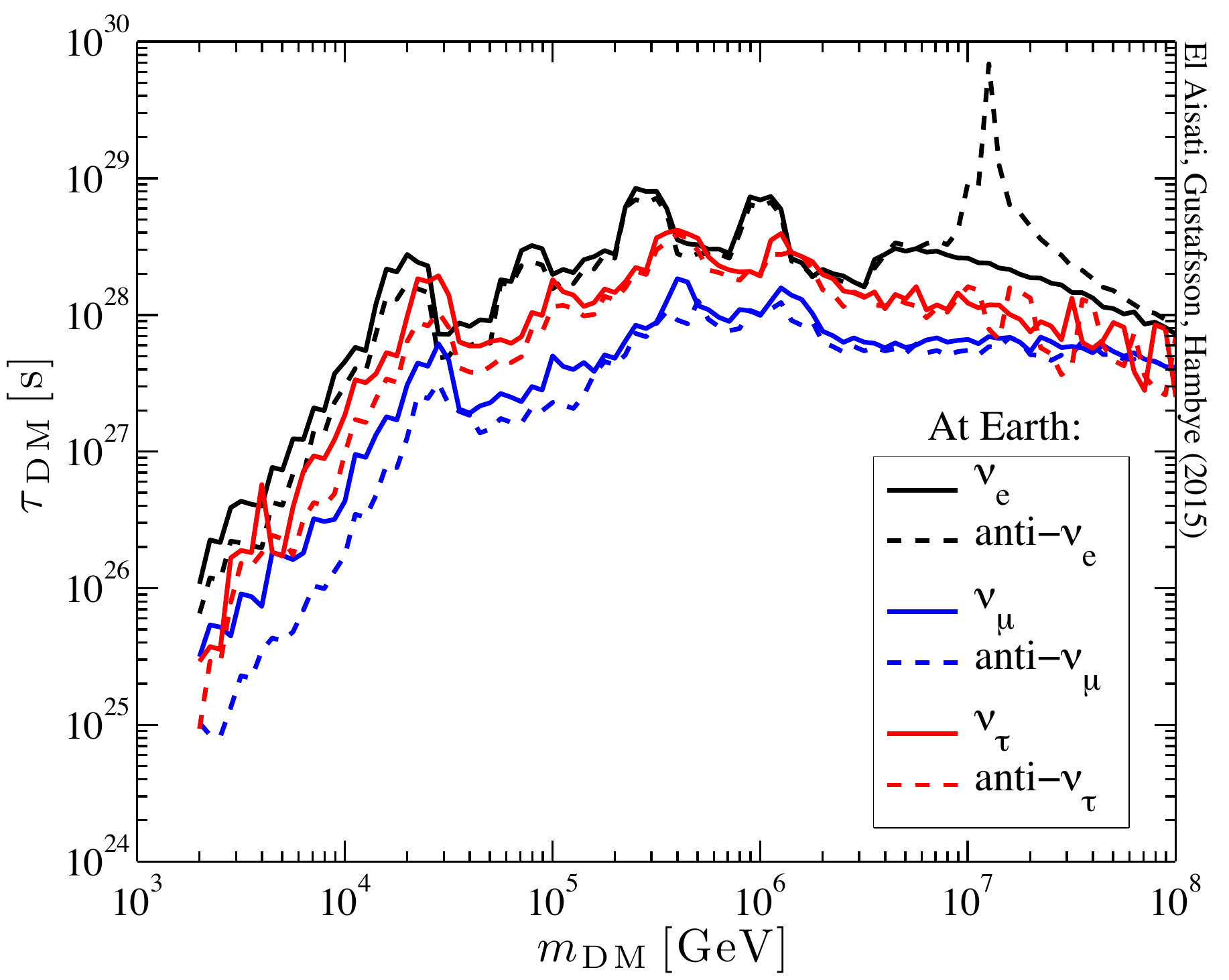}}
\caption{
Same as Fig.~\ref{fig:Source_anti} but in the case of pure $\nu_e$, $\nu_\mu$ and $\nu_\tau$ fluxes at {\it  Earth's surface}. A pure flavor flux is typically not realistic due to neutrino oscillations, but the plot clearly illustrates the differences in effective area depending on the incoming neutrino flavor.}
\label{fig:Earth_anti}
\end{figure}

For Fig.~\ref{fig:Earth_anti}, we assume that the incoming fluxes {\it at Earth} are 100\,\% pure $\nu_e$, $\nu_\mu$ or $\nu_\tau$. As in the previous figure,  we show limits separately for neutrinos and anti-neutrino injected at DM decays. These (hypothetical) pure flavor fluxes incoming to Earth could potentially be partly mimicked by non standard flavor oscillation effects from: neutrino decay \cite{Beacom:2002vi,Baerwald:2012kc}, sterile neutrinos \cite{Athar:2000yw}, a pseudo-Dirac nature of neutrinos \cite{Beacom:2003eu}, Lorentz or CPT violations \cite{Hooper:2005jp} or gravity-induced decoherence \cite{Anchordoqui:2005gj}. In any case, these limits clearly illustrate the main differences in sensitivity reach with this data set for extreme cases (i.e.\ for pure flavor and particle/anti-particle fluxes). Potentially, these limits could be further improved if flavor and anti-neutrino identification properties in the data was used (as is mentioned in Section~\ref{sec:improvements}). 

\section{Tabulated limits}
\label{sec:TableResult} 

Tables of all limits and TS values in this article can be found in the online supplementary material file \texttt{Appendix\_D\_Table.dat} [included among the source files submitted to the \texttt{arXiv} at http://arxiv.org].

\end{document}